\documentclass[twocolumn]{aastex631}
\usepackage[section]{placeins}
\usepackage{float}
\usepackage{amsmath}
\usepackage{bm}
\usepackage{hhline}
\usepackage{CJK}
\usepackage{xcolor}

\usepackage[normalem]{ulem}

\shorttitle{Machine-learning Cosmology from voids}
\shortauthors{Wang et al.}

\graphicspath{{./}{Results/}}

\begin{document}
\begin{CJK*}{UTF8}{bsmi}
\title{Machine-learning cosmology from void properties}

\correspondingauthor{Bonny Y. Wang}
\email{ywang@flatironinstitute.org}

\author[0000-0001-7168-8517]{Bonny Y. Wang (汪玥)}

\affiliation{The Cooper Union, 30 Cooper Sq, New York, NY 10003 USA}
\affiliation{Center for Computational Astrophysics, Flatiron Institute, 162 5th Avenue, New York, NY 10010 USA}

\author[0000-0002-6146-4437]{Alice Pisani}
\affiliation{The Cooper Union, 30 Cooper Sq, New York, NY 10003 USA}
\affiliation{Center for Computational Astrophysics, Flatiron Institute, 162 5th Avenue, New York, NY 10010 USA}
\affiliation{Department of Astrophysical Sciences, Princeton University, 4 Ivy Lane, Princeton, NJ 08544 USA}

\author[0000-0002-4816-0455]{Francisco Villaescusa-Navarro}
\affiliation{Center for Computational Astrophysics, Flatiron Institute, 162 5th Avenue, New York, NY 10010 USA}
\affiliation{Department of Astrophysical Sciences, Princeton University, 4 Ivy Lane, Princeton, NJ 08544 USA}

\author[0000-0002-5854-8269]{Benjamin D. Wandelt}
\affiliation{Center for Computational Astrophysics, Flatiron Institute, 162 5th Avenue, New York, NY 10010 USA}
\affiliation{Sorbonne Universit\'e, CNRS, UMR 7095, Institut d'Astrophysique de Paris, 98 bis bd Arago, 75014 Paris, France}

\begin{abstract}
Cosmic voids are the largest and most underdense structures in the Universe. Their properties have been shown to encode precious information about the laws and constituents of the Universe. We show that machine learning techniques can unlock the information in void features for cosmological parameter inference. We rely on thousands of void catalogs from the {\tt\string GIGANTES} dataset, where every catalog contains an average of 11,000 voids from a volume of $1~(h^{-1}{\rm Gpc})^3$. We focus on three properties of cosmic voids: ellipticity, density contrast, and radius. We train 1) fully connected neural networks on histograms from individual void properties and 2) deep sets from void catalogs, to perform likelihood-free inference on the value of cosmological parameters. We find that our best models are able to constrain the value of $\Omega_{\rm m}$, $\sigma_8$, and $n_s$ with mean relative errors of $10\%$, $4\%$, and $3\%$, respectively, without using any spatial information from the void catalogs.  Our results provide an illustration for the use of machine learning to constrain cosmology with voids.
\end{abstract}


\section{Introduction} \label{sec:intro}
Cosmic voids, the underdense regions in the galaxy distribution, are dominated by dark energy and account for most of the volume of the Universe \citep{Gregory1978,Joeveer1978,Kirshner1981, Zeldovich1982, Waygaert1993, Bond1996, Tikhonov2006}. Thanks to their underdense feature, voids are particularly sensitive to cosmological information \citep{Park2007, Lavaux2010,Bos2012, Lavaux2012, Pisani2015, Hamaus2016, Mao2017,Pisani2019, Sahlen2019, Verza2019,Ronconi2019,Chan2019,Bayer_2021,Kreisch2021, Wilson2021, Contarini2022,Pelliciari2022,Contarini2022b}. Until recently, void numbers were relatively low given the fact that voids are large regions and that the volume of surveys was relatively small. Nowadays, large-scale surveys are expected to enable big data approaches for studying voids, revealing their relationship to cosmological models in all its strength. Upcoming surveys will provide of the order of $10^5$ voids each, gaining more than an order of magnitude in void numbers \citep{Pisani2019,Moresco2022}. Along with the large increase in available data from current and upcoming surveys, void data from simulations is also dramatically increasing (see e.g. the extensive {\tt\string GIGANTES} set of void catalogs in \cite{Kreisch2021})---enabling the usage of machine learning to analyze voids and find correlations with the value of cosmological parameters.

To extract cosmological information from voids, robust theoretical predictions are necessary for each of the different void statistics; examples include the void size function, providing the void number density as a function of void radii, and the void-galaxy cross-correlation function, corresponding to the density profile of voids \citep[see e.g.][]{Hamaus2016,Verza2019, Pisani2019,Hamaus2020,Contarini2022}. Additionally, relating theoretical predictions with measurements from observations is not always easy, since models are constructed by assuming ideal, isolated voids in the theoretical setup, different from voids measured in observations \citep{Contarini2021,Stopyra2021}. Therefore, the current extraction of cosmological information from voids is limited by the current progress in modelling void statistics. While most of the modelling effort is geared towards traditional void statistics such as the ones mentioned above, the relationship between other void properties and cosmological parameters needs further exploration. Machine learning provides a well-established framework to perform this task. 

In the future, as the volume and density of sky surveys and simulations continue to grow, and more machine learning methods are adopted into this field, we believe that void statistics will stand as an increasingly effective and competitive probe for cosmology \citep{Ntampaka2019,Baron2019,Lemos2022}.

Two void properties deserve special attention: 1) the ellipticity and 2) the density contrast \citep{Lavaux2010, Biswas2010, Sutter2014, Sutter2015,Kreisch2021,Schuster2022}. In this work, we intend to use them, plus the traditional void radius, to train neural networks to perform likelihood-free inference on the value of the cosmological parameters. We now discuss why these properties may carry cosmological information.

A common expectation of voids is that they should have spherical shapes \citep{Icke1984,Dubinski1993,Waygaert1993}. In reality, due to the counterbalance between the tidal effect of the dark matter and the expansion of the Universe, voids are in fact far from spherically symmetric on a one-to-one basis \citep{Sheth2004,Shandarin2006,Park2007}. Previous work shows that the void ellipticity distribution depends quite sensitively on cosmological parameters and is a promising probe for constraining the dark energy equation of state \citep{Lavaux2010,Biswas2010, Bos2012}. In particular, \cite{Biswas2010} considered the void ellipticity distribution as a precision probe for dark energy parameters $\{w_0, w_a\}$ using the Fisher forecasts framework. 
In this work, we instead focus on predicting $\Omega_m$ (in a flat Universe corresponding to constraining $\Omega_\Lambda=1-\Omega_m$). Also, \cite{Sutter2014, Sutter2014b} investigated the match between analytical models of void ellipticity with simulations and observations. Other factors affect void ellipticities, including (but not limited to) the presence of redshift space distortions \citep{Shoji2012, Bos2012, Hamaus2020}, the possibility of dark matter and dark energy interactions \citep{Rezaei2020}, and modified gravity \citep{Perico2019, Zivick2015}. Despite the strong evidence of the sensitivity of void ellipticity to cosmology, there are still very limited studies aiming at using this property for cosmology. 

A similar statement can be made for the density contrast of voids, a measure of voids' depth with respect to their edges, formally defined for popular Voronoi-based void finders as the ratio of the minimum density of the particle on the ridge of the void to the minimum density of the void \citep{Neyrinck2008}. From a purely theoretical perspective, the void definition is linked to density contrast: voids can be defined as matter density fluctuations for which the mean density contrast in a sphere reaches a threshold value for an effective radius value. In void finding algorithms, density contrast has been a crucial factor for determining the probability of finding a void \citep[e.g.][]{Neyrinck2008,Sutter2015,Hoyle2002, Padilla2005}, and studies show that the void density contrast can be a relevant quantity to distinguish and remove spurious voids in a sample \citep{Cousinou2019}. The void density contrast is linked to the void density profile, an observable that has been deeply investigated in previous works \citep[e.g.][]{Lavaux2012, Sutter2012, Hamaus2014, Schuster2022}. So far, however, only few papers focus on directly utilizing the void density contrast as a stand-alone quantity to extract cosmological information.

In this paper, we attempt to answer the question of how much information is contained in these void properties. Given the expected complex and likely unknown likelihood of the data, we tackle this problem using machine learning. In particular, we aim to constrain the cosmological parameters $\{\Omega_m, \Omega_b, h, n_s, \sigma_8\}$ from either histograms of void properties (e.g. the distribution of void ellipticities) or from void catalogs directly. With respect to standard statistics (e.g. the power spectrum), void properties may have different degeneracies when inferring the value of parameters, so their usage may help break those degeneracies and get tighter constraints \citep{Bayer_2021,Kreisch2021,Pelliciari2022,Contarini2022b}.

In this paper, we build on previous works that have quantified the amount of information contained in the void size function \citep{Kreisch2021} and on void catalogs directly \citep{Cranmer2021} using machine learning techniques. \cite{Kreisch2021} carried out an initial proof-of-concept analysis, training neural networks to perform likelihood-free inference using the void size function. The void size function is one of the main void statistics expected to reach its golden era in the next few years \citep{Pisani2015,Pisani2019, Contarini2022}, with a theoretical model recently reaching a high maturity level \citep{Contarini2022b}. \citet{Cranmer2021} used deep sets to extract information from the three void properties we consider in this paper. However, in that work we were interested in developing new pooling operations and in making our model more interpretable. On top of that, we assumed a Gaussian posterior. In this work, we instead aim to maximize the extraction of information and perform likelihood-free inference. Furthermore, we analyze the different void properties more completely. 

This paper is organized as follows. Section \ref{sec:tool} describes the dataset and analysis tools used in this paper. It introduces the {\tt\string GIGANTES} dataset we employ, the algorithm for finding voids and the definitions for void properties (radius, ellipticity and density contrast). It also outlines the machine learning process used for our results. Section \ref{sec:analys} presents the results we obtained by training both fully connected neural networks on histograms from void properties and deep sets from void catalogs to perform likelihood-free inference of cosmological parameters. Finally, Section \ref{sec:conc} draws the main conclusions of this work, showing that void properties provide valuable information to constrain our model of the Universe.

\section{Methods}\label{sec:tool}
    \subsection{The {\tt\em\string GIGANTES} dataset}
        In this paper we train our models on data from {\tt\string GIGANTES}, a void catalog suite containing over a billion cosmic voids created by running the void finder {\tt\string VIDE} \citep{Sutter2015} on the spatial distribution of halos from the {\tt\string QUIJOTE} simulations \citep{Villaescusa-Navarro2020}. 
        {\tt\string VIDE} is a well-established public Voronoi-watershed void finder toolkit based on {\tt\string ZOBOV} \citep{Neyrinck2008}, and used in a variety of papers \citep[see e.g.][]{Hamaus2015,Hamaus2016, Hamaus2017,Hamaus2022, Baldi2016, Kreisch2021,Contarini2022,Doglass2022}. It is the primary tool used to create the {\tt\string GIGANTES} dataset. While {\tt\string VIDE} has a wide variety of features, one of the most important properties is its ability to capture shape, which has proven to be particularly effective to extract cosmological information (e.g. compared to a spherical void finder, see \citealp{Kreisch2021}). Different techniques can be used to model voids \citep[see e.g.][]{baddeley2015, Okabe2000, Colberg2008, Cautun2018}. Using the Voronoi tessellation and the watershed transform \citep{Platen2007}, {\tt\string VIDE} is able to identify voids in dark matter or tracer distributions and provides void information including void position (RA, Dec, redshift or x, y, z; depending on whether the catalog is from observations and light-cones, or from a simulation box), void radius (calculated as 
           \begin{equation}
            \label{eq:voidRadius}
            R = \left(\frac{3}{4\pi}V\right)^{1/3},
           \end{equation}
           where $V$ is the total volume of the Voronoi cells that the void contains), void ellipticity (see section \ref{sec:ellipAndDensDef}), and void density contrast (see section \ref{sec:ellipAndDensDef}), among others.
           
           {\tt\string GIGANTES} is the first dataset built for analyzing cosmic voids with machine learning techniques. It provides void catalogs for over 7,000 cosmologies, including fiducial and non-fiducial ones, and covers redshifts $z = 0.0, 0.5, 1.0, 2.0$, both in real space and in redshift space. Cosmological parameters $\{\Omega_m, \Omega_b, h, n_s, \sigma_8, M_v,w\}$ are varied throughout the simulations. 
           
           The dataset for our paper is constituted by the void catalogs from the high-resolution {\tt \string QUIJOTE} latin-hypercube simulations at $z=0.0$. Thus, we have 2,000 void catalogs covering a volume of $1~(h^{-1}{\rm Gpc})^3$ each, from 2,000 cosmological parameters arranged in a latin-hypercube with boundaries defined by
           \begin{eqnarray}
               \Omega_m &\in& [0.1, 0.5]\\
               \Omega_b &\in& [0.03, 0.07]\\
               h &\in& [0.5, 0.9]\\
               n_s &\in& [0.8, 1.2]\\
               \sigma_8 &\in& [0.6, 1.0].
           \end{eqnarray}
On average, each catalog contains about 11,000 voids. The large number of catalogs and voids allows us to train models to perform likelihood-free inference. In this work we quantify how much information is embedded into 1) the distribution of void ellipticities, 2) the distribution of void density contrasts, and 3) void catalogs that contain radius, ellipticity, and density contrast for each void. We now describe how the ellipticity and density contrast are computed for each void.
\subsection{Void ellipticity and density contrast}
        \label{sec:ellipAndDensDef} 
        \begin{figure}
            \centering
            \epsscale{1.1}
            \includegraphics[width=.45\textwidth]{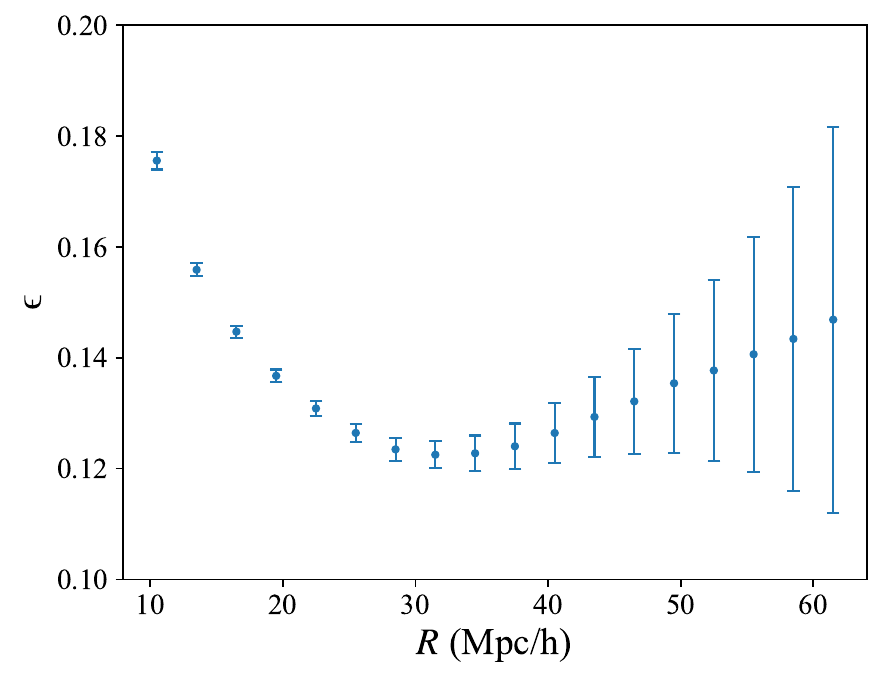}
            \includegraphics[width=.45\textwidth]{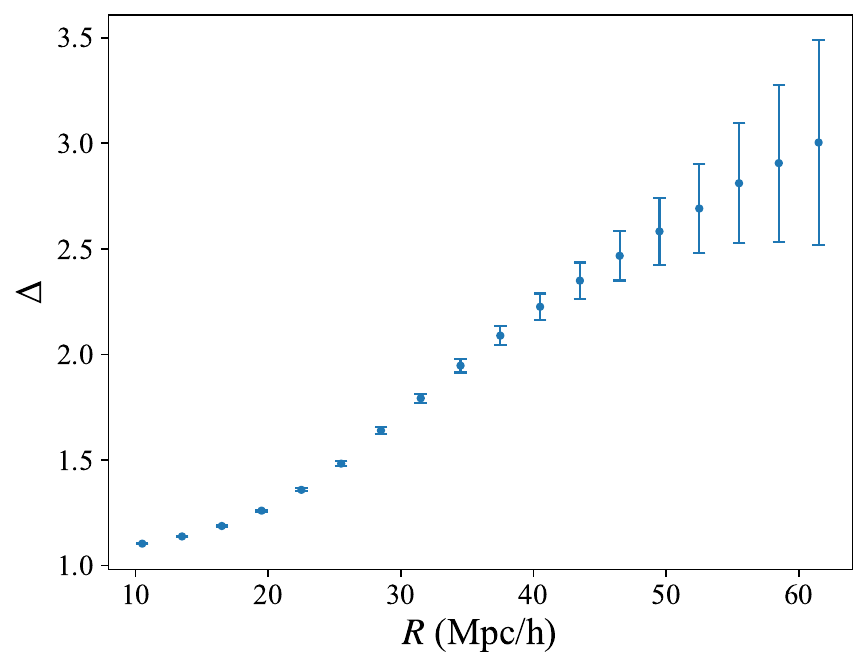}
    
            \caption{Void density contrast and ellipticity as a function of radii from all the simulations in the high-resolution {\tt \string QUIJOTE} latin-hypercube simulations at $z=0.0$. The blue dots and error bars represent the mean and the standard deviation in the mean of each bin. Void radii are binned from 10.5 Mpc/h to 61.5 Mpc/h.}
            \label{fig:radius_Relationship}
        \end{figure}
          The ellipticity of a void is computed by VIDE from the eigenvalues and eigenvectors of the inertia tensor \citep{Sutter2015} and defined as:
        \begin{equation}
            \label{eq:ellip}
            \epsilon = 1 - \left(\frac{J_1}{J_3}\right)^{1/4},
        \end{equation}
        where $J_1$ and $J_3$ are the smallest and largest eigenvalues of the inertia tensor, that is calculated as: 
        \begin{equation}
            \label{eq:inertia}
            \begin{split}
            & M_{xx} = \sum_{i = 1}^{N_p}(y_i^2 + z_i^2)\\
            & M_{xy} = -\sum_{i = 1}^{N_p}x_iy_i
            \end{split}
         \end{equation}
        and complemented with cyclic permutations. Here $N_p$ is the number of particles in the void while $x_i$, $y_i$, $z_i$ represent the cartesian coordinates of the particle $i$ with respect to the void macrocenter. The macrocenter in the {\tt\string VIDE} algorithm is defined as: 
        \begin{equation}
            \label{eq:macrocenter}
            \pmb{\mathrm{X}}_v = \frac{1}{\sum_i V_i} \sum_i \pmb{\mathrm{x}}_iV_i,
        \end{equation}
        where $\pmb{\mathrm{x}}_i$ and $V_i$ are the positions and Voronoi cell volumes of each particle $i$ in the void \citep{Sutter2015}.
        \begin{figure*}
            \centering
            \epsscale{1.2}
            \plotone{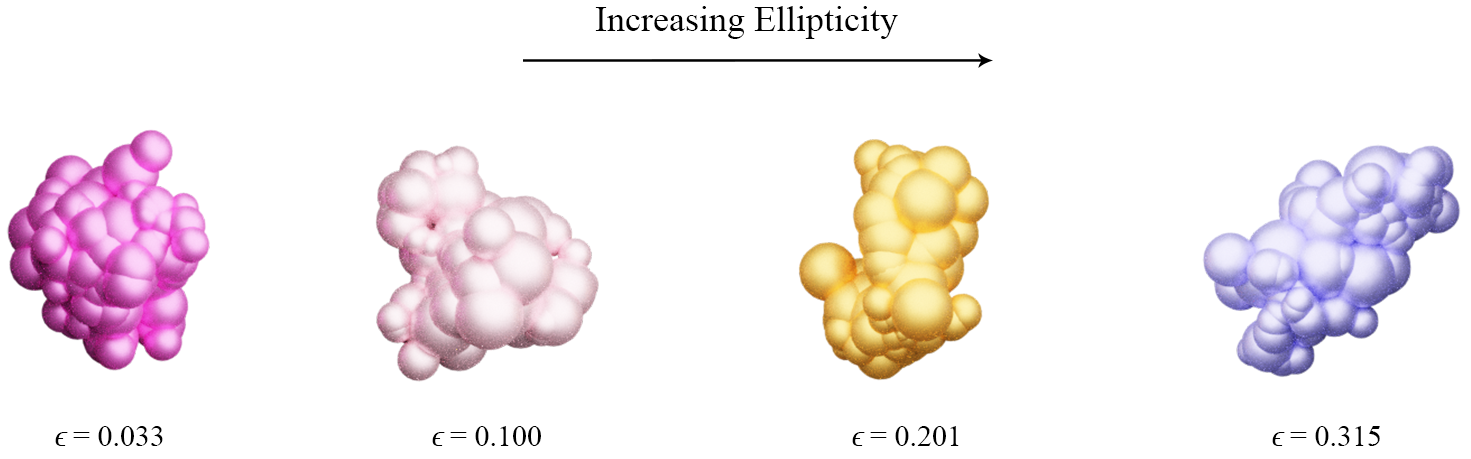}
            \caption{Examples of voids with different ellipticities. We use {\tt\string VIDE} to find the Voronoi cells of a void and plot the Voronoi cells of each void to show the void geometry. Small spheres inside each void represent its Voronoi cells. We show four randomly selected voids and arrange them by their ellipticities. We note that there is a large variety in the morphology of voids, so the voids shown in the figure should not be seen as representatives of their class. 
            }
            \label{fig:ellipVis}
        \end{figure*}
        \begin{figure*}[!htb]
            \centering
            \epsscale{1.24}
            \plotone{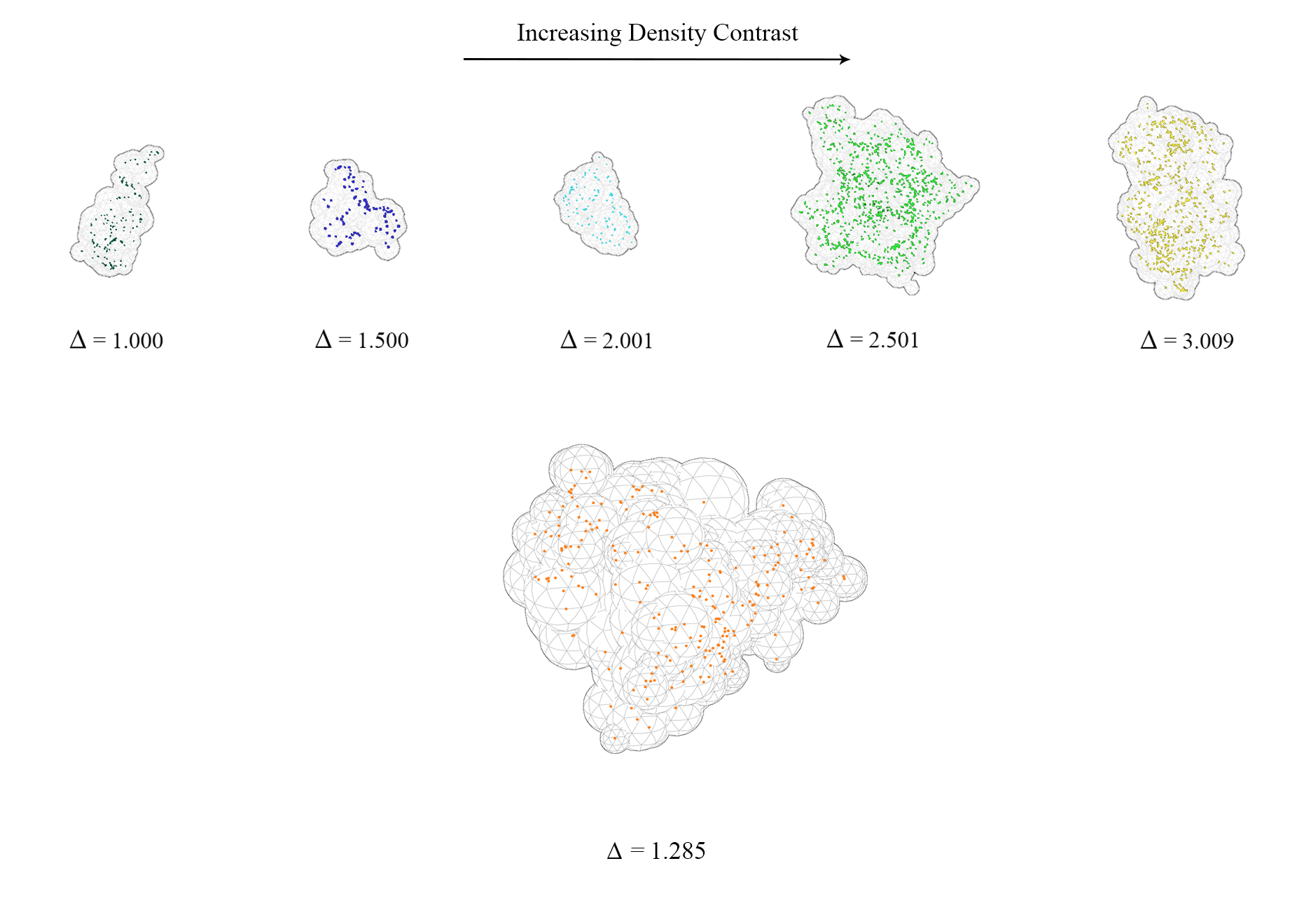}
          \caption{Examples of voids with different density contrasts. The gray wireframes show the geometries of the voids formed by the Voronoi cells of each void, similar to Figure \ref{fig:ellipVis}. The colored dots inside each void represent halos. Voids in the top row are arranged by increasing density contrast (from left to right). The larger void at the bottom allows a clearer view of a void with a density contrast of 1.285, the mean value in the catalogs.}
            \label{fig:dcVis}
        \end{figure*}
        The void density contrast\footnote{In \citealp{Sutter2015}, $r$ is used to represent void density contrast. However, to avoid confusion with radius, we use here $\Delta$ instead of $r$ for the void density contrast.}, $\Delta$, is defined\footnote{This definition for the density contrast is used by {\tt\string ZOBOV}, by {\tt\string VIDE} and, consequently in the {\tt\string GIGANTES} catalogs.} as the ratio of the minimum density along the ridge of the void, $\Delta_{ridge}$, versus the minimum density in the void, $\Delta_{min}$:
        \begin{equation}
            \label{eq:DC}
            \Delta = \frac{\Delta_{ridge}}{\Delta_{min}}.
        \end{equation}
        
        We show in Figure \ref{fig:radius_Relationship} the ellipticity and density contrast as a function of void radii. The void radii are binned from 10.5 Mpc/h to 61.5 Mpc/h, and the error bars are the standard deviations of the mean for each bin. For comparison purposes, in Figure \ref{fig:radius_Relationship} we use the same void radius range as in the {\tt\string GIGANTES} analysis \citep{Kreisch2021}, that is from 10.5 Mpc/h to 61.5 Mpc/h. As Figure \ref{fig:radius_Relationship} shows, when void radius is under 30 Mpc/h, void ellipticity decreases as the void radius increases. In this range, smaller voids are more elliptical than larger voids. This is also consistent with voids found in {\tt\string Magneticum} simulation \citep{Schuster2022} and voids in SDSS DR9 \citep{Sutter2014b}, and with the expectation that small voids are less evolved structures \citep{Sheth2004}. For voids with radii larger than 30 Mpc/h, we notice that an increase in radius leads to a mild increase in the mean values of ellipticity. However, this increase is accompanied by a notable expansion of the error bars, which can be attributed to the scarcity of voids within that range. The bottom panel shows that the void density contrast increases with void radius. This is likely related to the locations of the detected voids, since smaller voids are more commonly found in denser regions or by the edges of the larger voids, corresponding to local over-densities. Therefore, there might be smaller changes in matter density across smaller voids. On the other side, larger voids have usually been evolving for a longer time than small voids \citep{Sheth2004}. Therefore, even if dark energy generically has a late-time impact, larger voids were dominated by dark energy earlier than smaller voids. Consequently, dark energy within large voids has had more time to make those voids larger and consequently emptier, causing a greater density contrast.
Figures \ref{fig:ellipVis} and \ref{fig:dcVis} show examples from our dataset of voids with different ellipticities and density contrasts (respectively).

In our analysis we do not impose any cut to the voids properties (e.g. only considering voids with ellipticities smaller than a given value) since machine learning methods can learn to marginalize over features that are not correlated with cosmology \citep{Villaescusa2021}.
\subsection{Machine learning}

\begin{figure}
        \centering
        
        \includegraphics[scale = 0.3]{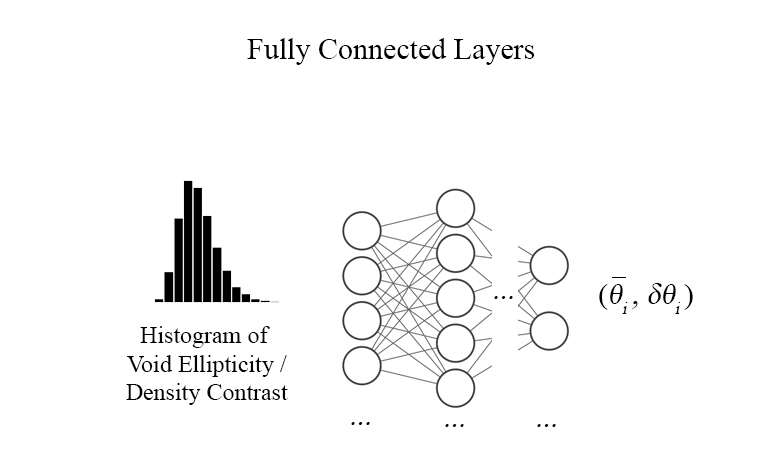}
        \includegraphics[scale = 0.3]{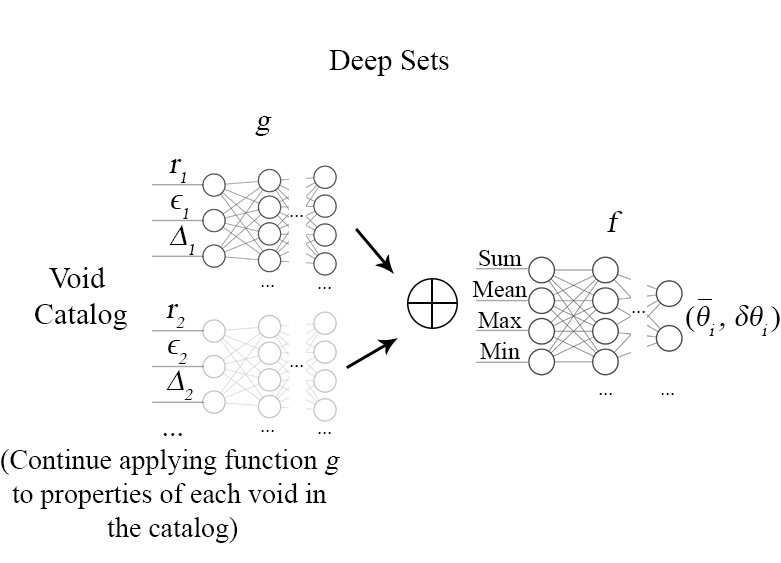}
        \caption{Visualizations of the two machine learning architectures used in this paper. While fully connected layers are one of the most commonly used neural network architectures, deep sets are able to extract information directly from the void catalog regardless of the number of voids in the catalog. Functions $f$ and $g$ in the deep sets architecture are modeled with fully connected layers. $\bigoplus$ represents a permutation invariant operation. More details about the deep sets architecture are given in section \ref{subsec:deep_sets}.}
        \label{fig:MLVis}
    \end{figure}

    The large {\tt\string GIGANTES} dataset enables cosmic void exploration with machine learning methodologies. In this\\
    \\
    \\
    work, we train neural networks to perform likelihood-free inference on the value of five cosmological parameters: ${\{\Omega_m, \Omega_b, h, n_s, \sigma_8\}}$.
    We make use of two different architectures that perform likelihood-free inference but whose input is different:
    \begin{itemize}
        \item \textbf{Fully connected layers.} The inputs to these models are 1-dimensional arrays. In our case, the data can be either the distribution of void ellipticities or the distribution of void density contrasts.
        \item \textbf{Deep sets.} The inputs to these models are void catalogs. The catalogs will contain a set of voids (each catalog can have a different number of voids) where each void will be characterized by three properties: 1) ellipticity, 2) density contrast, 3) radius.
    \end{itemize}
    In figure \ref{fig:MLVis}, we provide a visual explanation of these two architectures.
    
    The output of the two models is a 1-dimensional array with 10 values (2 values per parameter): 5 numbers for the posterior mean and 5 numbers for the posterior standard deviation. In order to train the networks to output these quantities, we use the loss function of moment networks \citep{Jeffrey2020}. To avoid problems that arise from different terms having different amplitudes, we use the change described in \citet{CMD}. Thus, the loss function we employ is: 
        \begin{eqnarray}
            \label{eq:loss}
            L = \frac{1}{|B|} \Biggl[ \log \left( \sum_{j \in B}\left(\theta_j - F(x_j)\right)^2\right)+\nonumber \\
             \log \left( \sum_{j \in B}\left(\left(\theta_j - F(x_j)\right)^2 - G(x_j)^2\right)^2 \right) \Biggr]~,
        \end{eqnarray}  
        where $F(x)$ and $G(x)$ are functions that will output the marginal posterior mean and standard deviation when input $x$ in batch $B$ for the true value of parameters $\theta$. The sum runs over all elements in a given batch and the functions are parametrized as fully connected layers or deep sets as described above. In the next section, we provide further details on the architecture and training procedure.
        
        In order to evaluate the accuracy and precision of the model, we make use of four different statistics: 1) the coefficient of determination $R^2$, 2) the root mean squared error, RMSE, 3) the mean magnitude of the relative error MMRE, and 4) the normalized chi-squared, defined as 
        \begin{equation}
            \chi^2=\frac{1}{N}\sum_{i=1}^N\left(\frac{\rm true-mean}{\rm standard\,deviation}\right)^2~,
        \end{equation}
        where mean and standard deviation represent the posterior mean and standard deviation predicted by the model, corresponding to $F(x)$ and $G(x)$ in equation \ref{eq:loss}. This measurement is directly related to the second term that we want to minimize in the loss function. The normalized chi-squared is useful to determine whether the errorbars are accurately determined: a value larger/smaller than 1 indicates that the errors are underestimated/overestimated. On the other hand, we mostly rely on the $R^2$ values to evaluate the linear correlations between our models' predictions and the true values.

\section{Results} \label{sec:analys}
\begin{figure*}[!htb]
    \includegraphics[width=.32\textwidth]{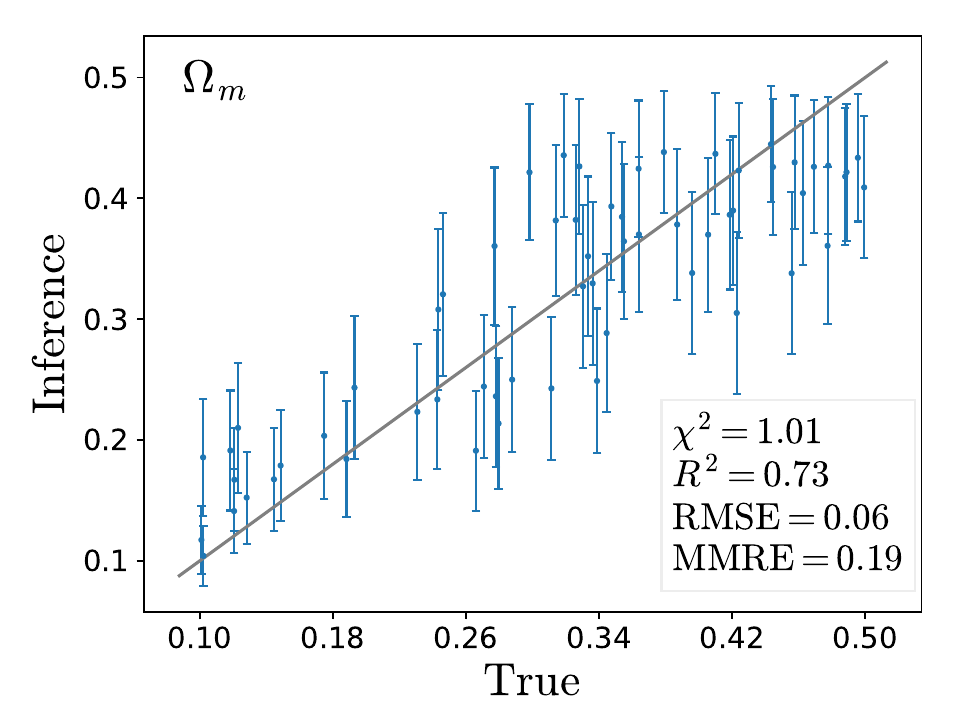}\\
    \includegraphics[width=.32\textwidth]{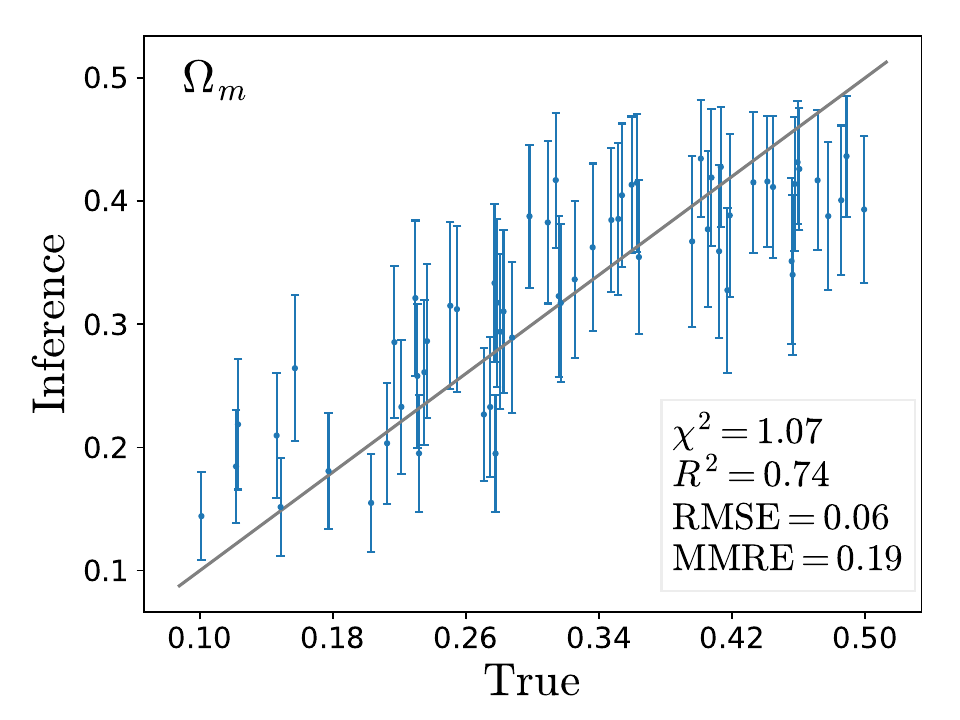}
    \includegraphics[width=.32\textwidth]{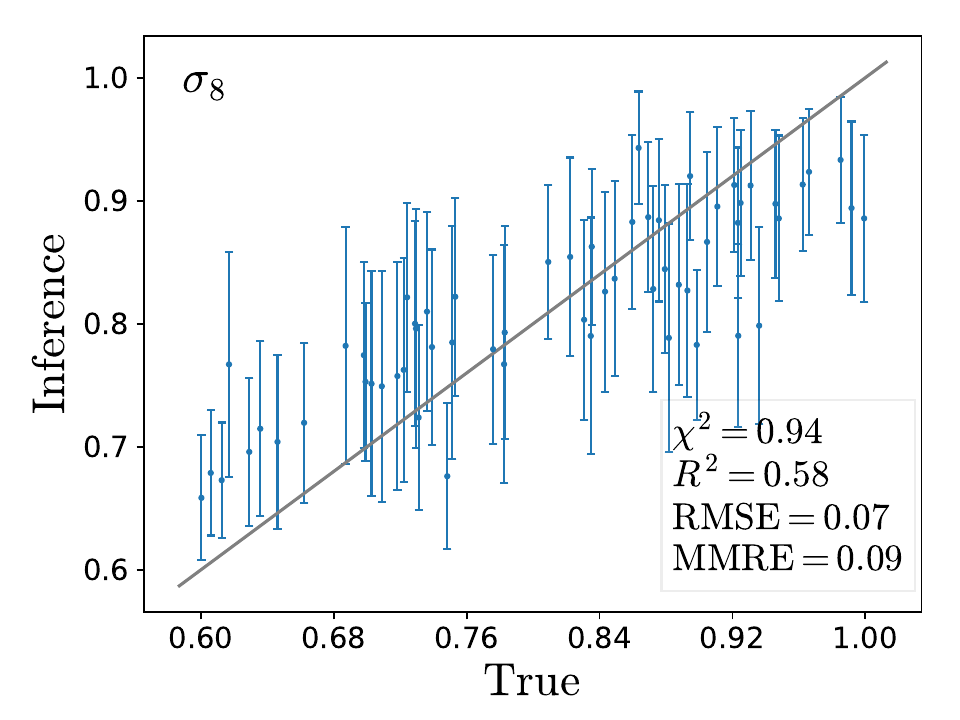}\\
    \includegraphics[width=.333\textwidth]{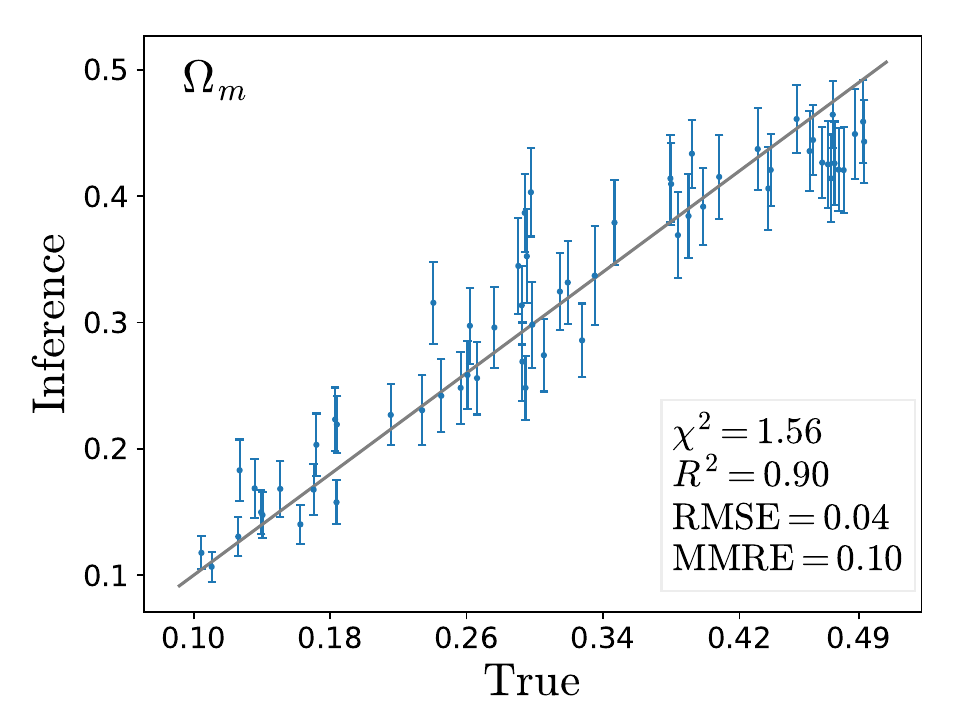}
    \includegraphics[width=.333\textwidth]{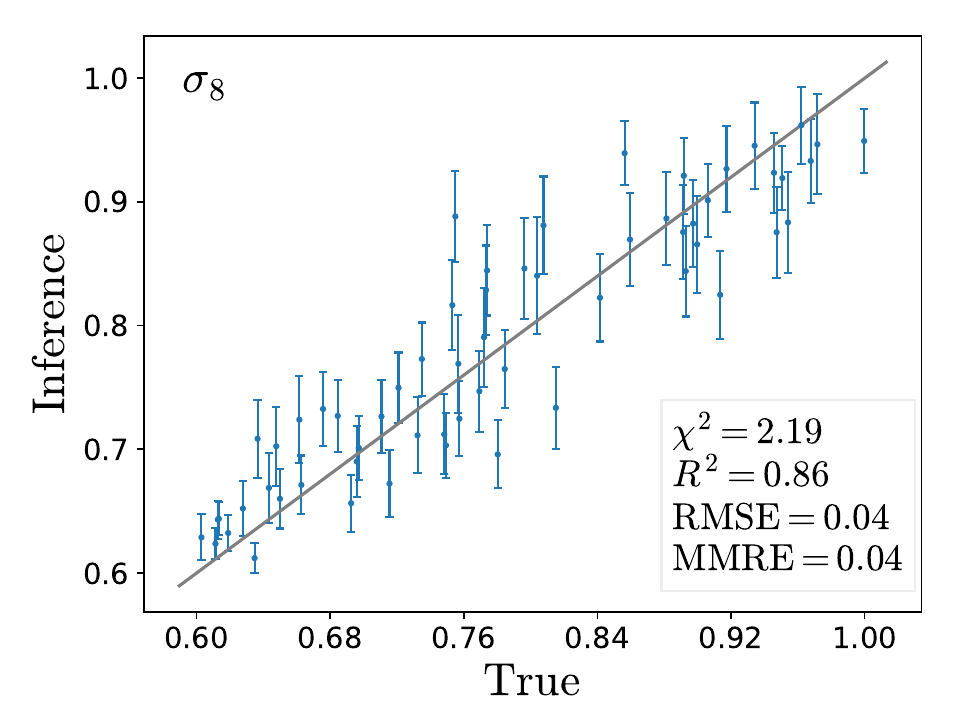}
    \includegraphics[width=.333\textwidth]{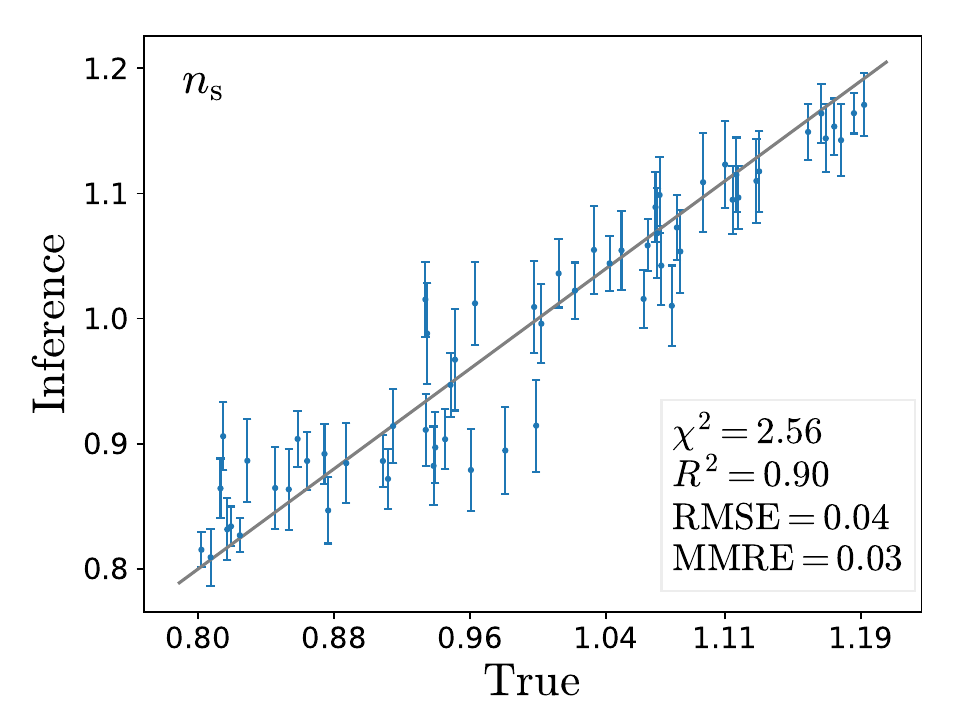}
    \caption{We train different models to perform likelihood-free inference on the value of the cosmological parameters. \textbf{Top row:} we use fully connected layers to infer the value of $\Omega_{\rm m}$ from histograms that describe the distribution of void ellipticities. \textbf{Middle row:} we train fully connected layers to infer the value of $\Omega_{\rm m}$ and $\sigma_8$ from the histograms that characterize the distribution of the density contrast of voids. \textbf{Bottom row:} we train deep sets on void catalogs to infer the value of $\Omega_{\rm m}$, $\sigma_8$, and $n_s$. Our model takes a void catalog  as input (containing all the voids in a given simulation), where every void is characterized by three properties: 1) ellipticity, 2) density contrast, and 3) radius. In all the cases, the models predict the posterior mean and standard deviation for each parameter. As expected, deep sets perform better than the histogram models. The bottom right part of the panels shows the accuracy metrics calculated from all the results obtained from the 198 values in the testing set. We find that the deep set model is able to infer the correct value of $\Omega_{\rm m}$ (left column), $\sigma_8$ (middle column), and $n_s$ (right column) with a mean relative error of $10\%$, $3\%$, and $4\%$, respectively.\\}
    \label{fig:deep_sets}
\end{figure*}
\begin{table*}
\centering
    \begin{tabular} {|>{\bfseries}
    p{3cm}| >{\centering} p{2cm}| >{\centering} p{2cm}|>{\centering} p{2cm}|>{\centering} p{2cm}|>{\centering} p{2cm}| wc{2cm}|}
    \hhline{|=======|}
    Type & \multicolumn{3} {c}{Histograms} & \multicolumn{3} {|c|}{Deep sets} \\
    \hline
    Void Property & Ellipticity & \multicolumn{2}{c}{Density contrast} & \multicolumn{3}{|c|}{ Ellipticity, density contrast, radius}\\
    \hline
    Parameter & $\Omega_m$ & $\Omega_m$ & $\sigma_8$ & $\Omega_m$ & $n_s$ & $\sigma_8$\\
    \hhline{|=|=|=|=|=|=|=|}
    $\chi^2$ & 1.01 & 1.07& 0.94 & 1.56& 2.56&2.19\\
    $R^2$ & 0.73& 0.74& 0.58& 0.90& 0.90& 0.86 \\
    ${\rm RMSE}$ & 0.06& 0.06& 0.07& 0.04& 0.04& 0.04\\
    ${\rm MMRE}$ & 0.19& 0.19& 0.09& 0.10& 0.03 & 0.04 \\
    \hhline{|=======|}
    \end{tabular}
\caption{Accuracy metrics on the considered cosmological parameters achieved from the various methods and datasets used in this work.}
\label{Tab:compare}
\end{table*}

    In this section, we show the results we obtain when inferring the value of cosmological parameters using the models trained with the different inputs. We split the data set into training(80\%), validation(10\%), and testing(10\%) set. We choose the best model of each architecture based on the model's performance on the validation set and show the results from the test sets. We first present the results for the histograms, and later we display the results of the deep sets. While the models are trained to infer the value of all five cosmological parameters, they may only be able to infer the value of some of them. In this case, we only display the results for the parameter(s) that can be inferred. We include in Appendix \ref{app:Unsucessful} the plots for cases when our models are not able to infer the parameters.
    
    All the errors we report correspond to aleatoric errors. We have also estimated the epistemic errors (associated to the error from the network itself) by training 10 different models with the best value of the hyperparameters and computing the standard deviation of the posterior means. We find the epistemic errors to be significantly smaller than the aleatoric ones: for the histograms, they are about 1/7 of the aleatoric ones while for the deep sets the ratio is 1/3. Since the error budget is dominated by the aleatoric errors, we do not report the value of the epistemic errors.
    \subsection{Ellipticity histograms}
    \label{sec:his}
    
    We start by training fully connected layers on ellipticity histograms that are constructed as follows. First, all the voids in a given simulation are selected. Second, the void ellipticities are assigned to 18 equally spaced bins from $\epsilon = 0.0$ to $\epsilon = 1.0$. Next, the number count in each bin is divided by the total number of voids in the considered simulations. In this way, we construct 2,000 ellipticity histograms, one per simulation. We further preprocess the data so that each bin has a mean of 0.0 and a standard deviation of 1.0.
    
    The architecture of the model consists of a series of blocks that contain a fully connected layer with a LeakyRelu non-linear activation function and a dropout layer. We note that the last layer only contains a fully connected layer. To find the hyperparameters for our model, we use the TPESampler in Optuna package \citep{Akiba2019} to perform Bayesian optimization over the following ranges of each hyperparameter: \begin{itemize}
        
        \item Number of hidden layers $\in [1,5]$
        \item Number of neurons in each hidden layer $\in[4,1000]$
        \item Learning rate $ \in [0.1, 1.0 \times 10^{-5}]$
        \item Weight decay $ \in [1.0, 1.0 \times 10^{-8}]$
        \item Dropout rate $\in[0.2, 0.8]$
    \end{itemize}
    The optimization is carried out by minimizing the validation loss for a total of 1,000 trials. For each trial, the training is performed through gradient descent using the AdamW optimizer \citep{Loshchilov2017} for 1,000 epochs and a batch size of 128. 
    
    We find that the model trained on ellipticity histograms can only infer the value of $\Omega_{\rm m}$, and we show the results of inferring the value of this parameter from the ellipticity histograms of the test set in the top row of Figure \ref{fig:deep_sets}. We find that the model is able to infer the value of $\Omega_{\rm m}$ with an RMSE value of 0.06 and a mean relative error of $19\%$. We report the values of the four accuracy metrics in Table \ref{Tab:compare}.

    \subsection{Density contrast histograms}
    
    We also train fully connected layers on density contrast histograms, that are constructed and standardized in the same way as the ellipticity histograms. They contain 18 bins equally spaced from $\Delta = 1.0$ to $\Delta = 3.0$. The model architecture, training procedure, and hyperparameter tuning are the same as the ones outlined above.
    
    In this case, we find that the model is able to infer both $\Omega_{\rm m}$ and $\sigma_8$. We show the results of this model in the middle row of Figure \ref{fig:deep_sets}. For $\Omega_{\rm m}$, the accuracy and precision of the model are comparable to the one achieved by the model trained on ellipticity histograms. For $\sigma_8$ the model can infer its value with a mean relative error of $9\%$. From the values of the $\chi^2$, we can conclude that the magnitudes of the error bars are well estimated. Looking at the value of the $R^2$, we can see that the estimate of $\Omega_{\rm m}$ is more accurate than the estimate of $\sigma_8$. We report all the accuracy metrics in Table \ref{Tab:compare}. 
    
    We have also trained neural networks using both ellipticity and density contrast histograms. In this case, the input to the network is a 1-dimensional vector with 36 dimensions: 18 from the ellipticity histogram and 18 from the density contrast histogram. Using this setup, we find similar results as the ones we obtain using the density contrast histograms, possibly indicating that the information in the two statistics may be somewhat redundant.
     
\subsection{Deep sets}
\label{subsec:deep_sets}
We now illustrate a different architecture to extract information from the void catalogs directly. In this work, we are not quantifying the cosmological information encoded in the clustering of voids, but just on their intrinsic properties. For this reason, we make use of deep sets \citep{DeepSets}: an architecture that is able to work with sets of objects whose number may vary from set to set. Each object in the set has a collection of properties associated to it, $\vec{p}$. Our goal is to train a model on void catalogs, to perform likelihood-free inference on the value of the cosmological parameters $\vec{\theta}$. 

Our model will take all the voids in a given simulation  as input and output the posterior mean ($\bar{\theta_i}$) and standard deviation ($\delta\theta_i$) of the cosmological parameter $i$. Every void is characterized by three numbers: 1) its ellipticity, 2) its density contrast, and 3) its radius. We relate the input and the output through:
\begin{equation}
(\bar{\theta_i}, \delta\theta_i) = f\left(\bigoplus g(\vec{p})\right)
\end{equation}
where $f$ and $g$ are neural networks (in this case fully connected layers) and $\bigoplus$ is a permutation invariant operation that applies to all the elements in the deep sets. In our case, we use four different permutation invariant operations: 1) the sum, 2) the mean, 3) the maximum, and 4) the minimum. The result of these four operations is concatenated before being passed to the last fully connected layer $f$. 

The neural networks $g$ and $f$ are parametrized as three and four fully connected layers where the number of neurons in the latent space is considered as a hyperparameter. We use the LeakyRelu activation functions with slope -0.2 in all layers but the last. We train the model for 2000 epochs with a batch size of 6 using the same loss function in Equation \ref{eq:loss} in order to perform likelihood-free inference. 

Similar to the hyperparameter optimization process of the fully connected layers, we use Optuna to perform Bayesian optimization for the deep sets model. The ranges of the hyperparameters are defined below:
\begin{itemize}
    \item Number of neurons in each hidden layer $\in[10,1000]$
    \item Learning rate $ \in [1.0 \times 10^{-4}, 1.0 \times 10^{-8}]$
    \item Weight decay $ \in [1.0 \times 10^{-3}, 1.0 \times 10^{-12}]$
\end{itemize}Here, we only perform 50 trials to optimize hyperparameters because of the longer training time. Optuna is run to minimize the value of the validation loss. While we train the model to predict the value of the five cosmological parameters we consider, we find that we are only able to infer the value of $\Omega_{\rm m}$, $n_s$, and $\sigma_8$. We show the results in Figure \ref{fig:deep_sets}.

Our model is able to infer the value of $\Omega_{\rm m}$, $n_s$, and $\sigma_8$ with a mean relative error of approximately $10\%$, $3\%$, and $4\%$, respectively. This represents a significant improvement over the results we obtained using histograms of the void ellipticities or density contrasts, and is also reflected in the values of the $R^2$ statistics. 
This is of course expected, given that with a sufficiently complex model, this method will not throw away information when we input the data to the networks\footnote{When we use histograms, we are losing information as we can only sample the distribution with a finite number of bins.}. We note that for $n_s$ and $\sigma_8$, our model achieves a slightly high value of $\chi^2$. We have checked that this is mainly driven by a single void catalog that has a slightly extreme cosmology. If we had removed this catalog, the obtained $\chi^2$ would be much closer to 1. Thus, we think our error bars are not strongly underestimated.

The reason for unsuccessful predictions could simply be the lack of information in the datasets we are using. However, we note that the parameter $\Omega_b$ will only change the amplitude and shape of the initial power spectrum in N-body simulations. Other works have shown that this effect is smaller than the ones induced by other parameters such as $\Omega_m$ and $\sigma_8$ (see e.g. \citet{Uhlemann2020}).

We note that these results could be further improved if more void properties were to be used, or if the network would exploit clustering information \citep{Pablo_2022, Lucas_2022, Shao_2022}.
\section{Conclusions}
\label{sec:conc}
In this work, we have shown how to connect void properties to the value of cosmological parameters through machine learning. We have trained several models to perform likelihood-free inference on the value of cosmological parameters from several void properties. We find that each individual property---ellipticity and density contrast---is sensitive to cosmological parameters, and that their combination with void radius yields the tightest constraints.

Currently, void ellipticity has not been used to extract constraints from data, due to the lack of robust modeling for the considered void properties. However other void statistics have been used to provide constraints from data that showcase the constraining power of voids. The void-galaxy cross-correlation function has been used to constrain $\Omega_m$ \citep{Hamaus2020}, finding a value of 0.310 with a relative error of 5\% (compared to the reference value $0.315 \pm 2\%$ in \citet{Plank2018}). This work relies on data from the BOSS survey \citep{Shadab2017}, whose volume is much larger than the volume of the simulation box we used in this work---it is therefore expected that we have a larger error. The void size function has also been used to constrain cosmological parameters in the BOSS survey, obtaining $\Omega_m = 0.29 ^{+24\%}_{-20\%}$ and $\sigma_8 = 0.80^{+11\%}_{-10\%}$ constraints \citep{Contarini2022c}.

We note that there are several caveats in this work. First, we have identified voids from the spatial distribution of dark matter halos in real space. A more realistic exercise will be to find voids from galaxies in redshift space. Second, we have assumed that the considered void properties are robust to uncertainties from N-body codes and hydrodynamic effects. While dark matter halo positions and velocities have been found to be robust to these effects \citep{Shao_2022}, the same exercise needs to be performed for the void properties considered in this work \citep[see also][]{Schuster2022}. Third, we have neglected the effect of super-sample covariance in our analysis due to the finite volume of the {\tt\string QUIJOTE} simulations. However, a recent study has shown that the effect of super-sample covariance on the void size function is small \citep{Bayer_2022}.

We emphasize that in this work we have only attempted to extract information from a few void properties. Our constraints can be improved by using more void properties (e.g. the number of galaxies in a void, or the value of the external tidal tensor affecting the void), or by exploiting the clustering of voids \citep[e.g. by using graph neural networks, see ][]{Pablo_2022, Lucas_2022, Shao_2022}, which we leave for future work.

Additionally, in future work we plan to investigate if combining void properties with traditional clustering statistics (e.g. halo power spectrum) yield tighter constraints on the value of cosmological parameters or whether the cosmic void information is somehow redundant to what machine learning can extract from traditional statistics. Finally, the distribution of void ellipticity was proven to be sensitive to the dark energy equation of state parameters ($w_0$ and $w_a$) as well \citep{Biswas2010}: it would therefore be interesting to use the methods presented in this paper to constrain the dark energy equation of state parameters and compare constraints from machine learning to those from the Fisher forecasts in \cite{Biswas2010}. Since the dataset we use here does not vary $w_0$ and $w_a$, we leave this for future work.

This work sets the path for future research directions. We have shown that void properties encode information about the value of cosmological parameters, and that machine learning provides access to that information. If the challenges discussed above can be controlled, we could train these methods on voids from numerical simulations and test them on voids from large-scale surveys such as DESI \citep{DESI2016}, Euclid \citep{Laureijs2011}, Roman \citep{Spergel2015}, SPHEREx \citep{Dore2018}, and PFS \citep{Tamura2016}, in order to extract cosmological information about our Universe.

\section{acknowledgments}
We thank Allister Liu, Samuel Keene, Wenhan Zhou, and Carolina Cuesta-Lazaro for helpful discussions and suggestions for this work. The Center for Computational Astrophysics at the Flatiron Institute is supported by the Simons Foundation.


\appendix
\section{Unsuccessful Inferences}
\label{app:Unsucessful}
\begin{figure*}[!htb]
    \includegraphics[width=.333\textwidth]{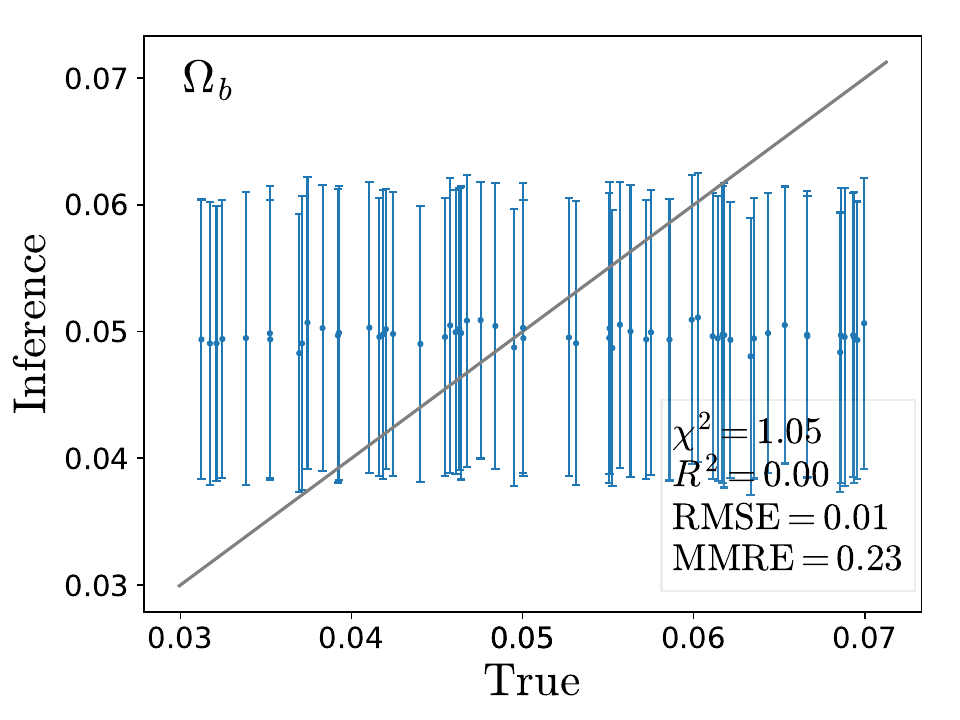}
    \includegraphics[width=.333\textwidth]{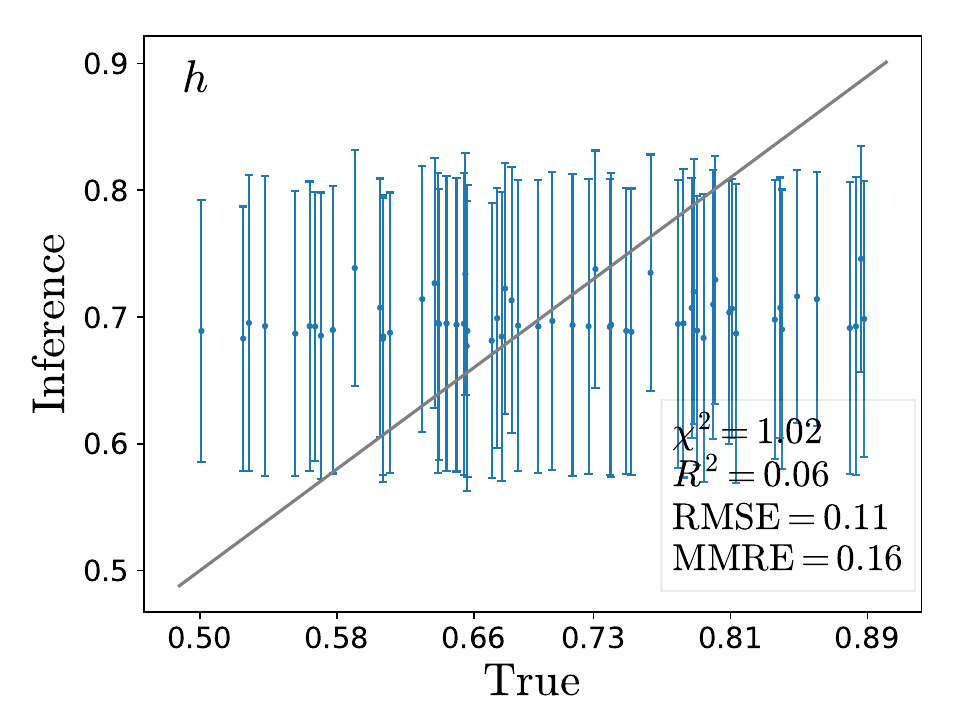}
    \includegraphics[width=.333\textwidth]{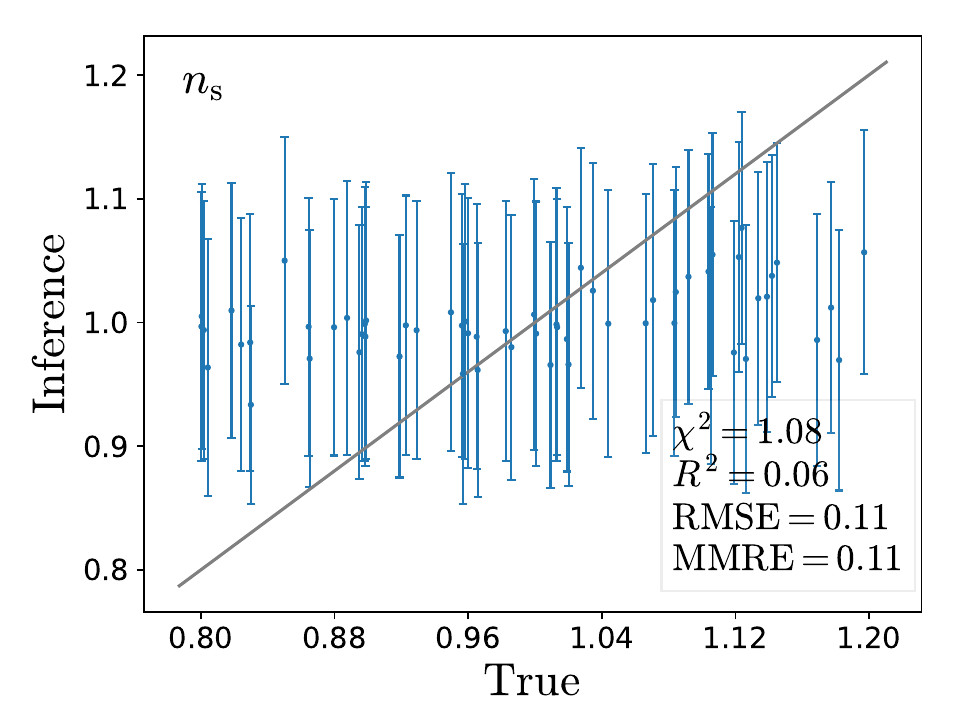}

    \caption{Plots representing the results for  the unconstrained parameters using fully connected layers with density contrast histograms.}
    \label{fig:my_label}
\end{figure*}
\begin{figure*}[!htb]
     \includegraphics[width=.333\textwidth]{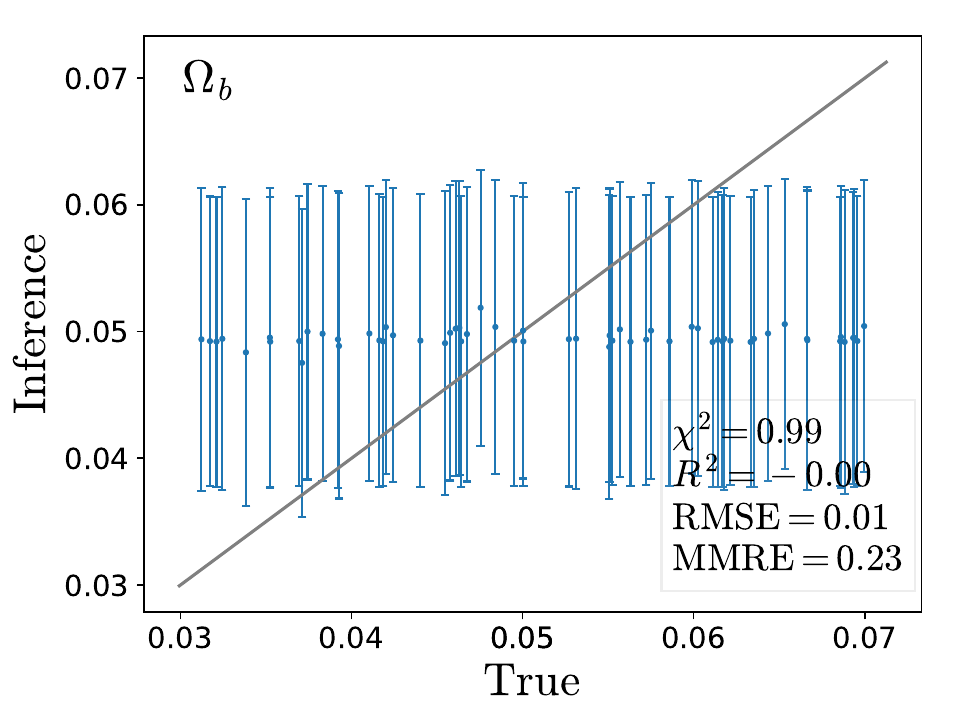}
    \includegraphics[width=.333\textwidth]{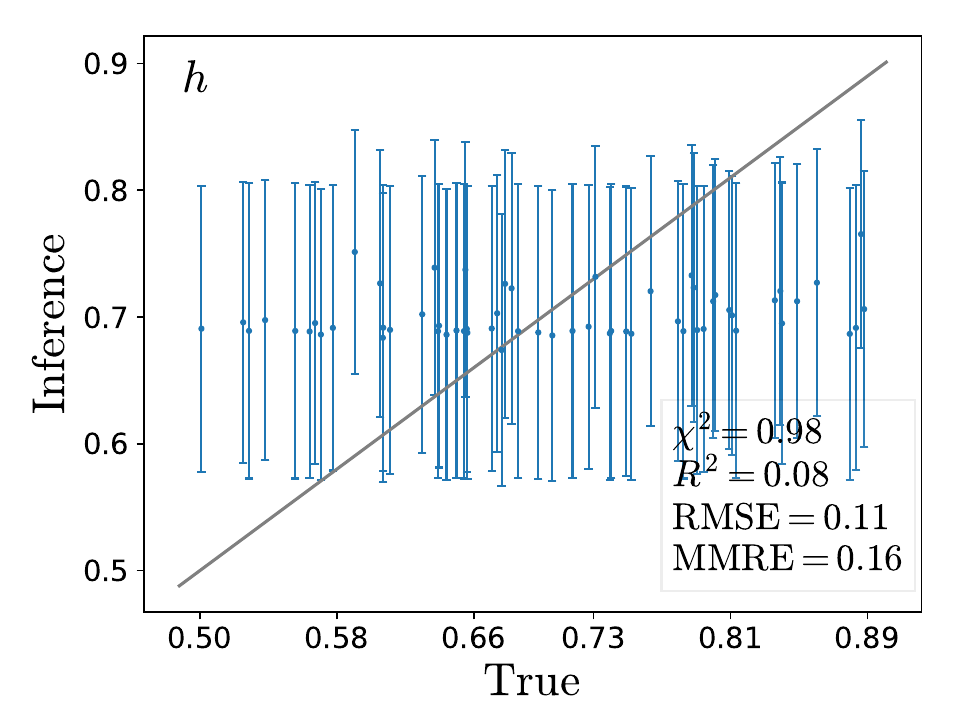}
    \includegraphics[width=.333\textwidth]{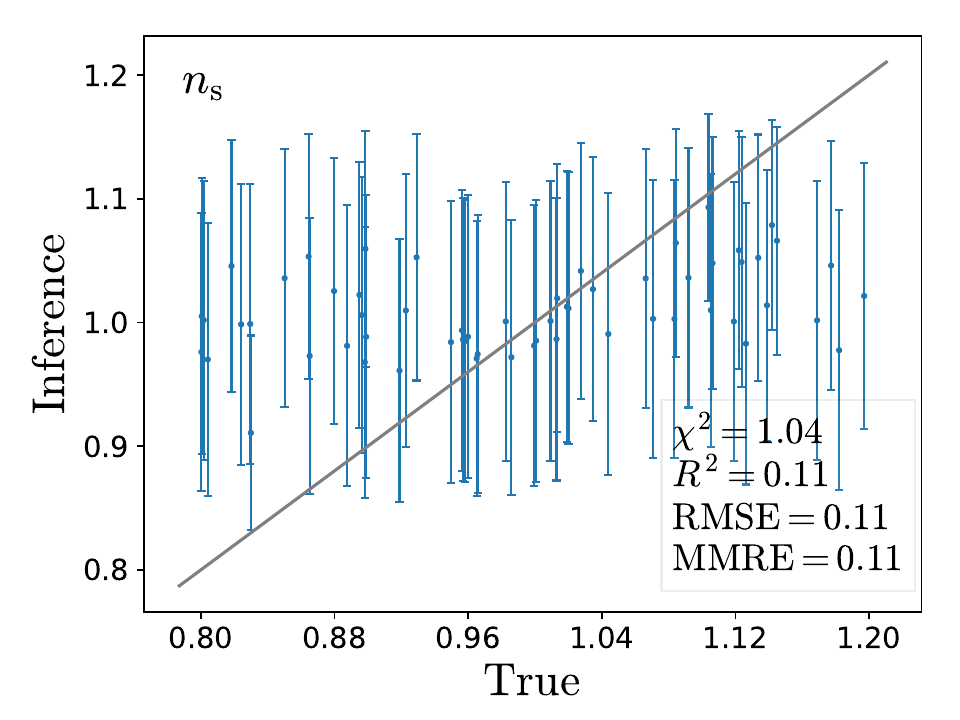}

    \includegraphics[width=.333\textwidth]{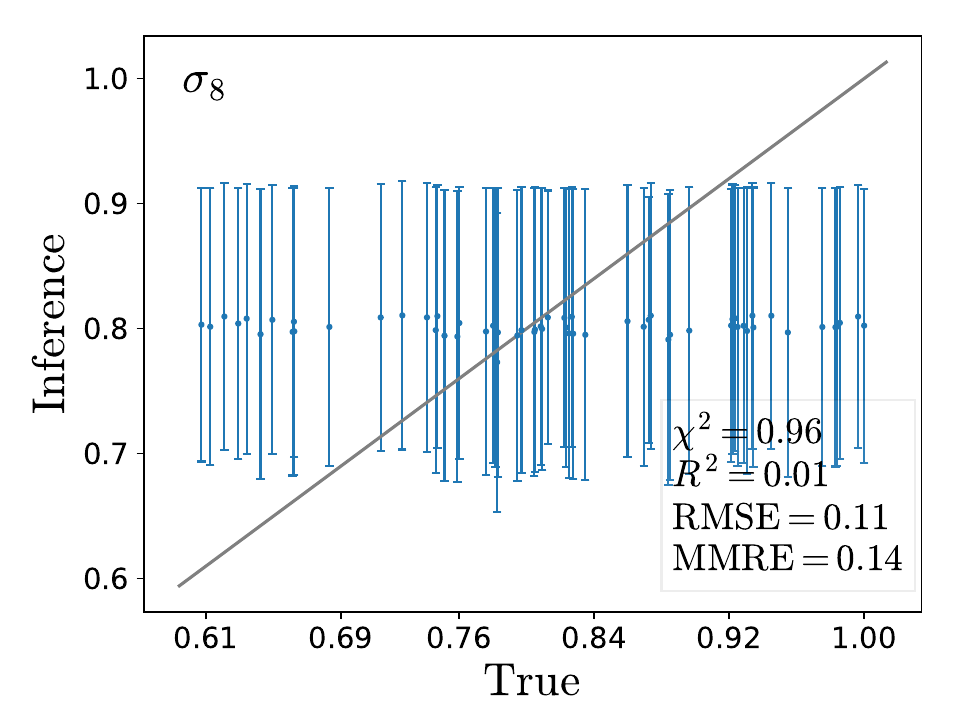}
  \caption{Plots representing the results for unconstrained parameters using fully connected layers with ellipticity histograms.}
\end{figure*}

\begin{figure*}[!htb]
    \includegraphics[width=.333\textwidth]{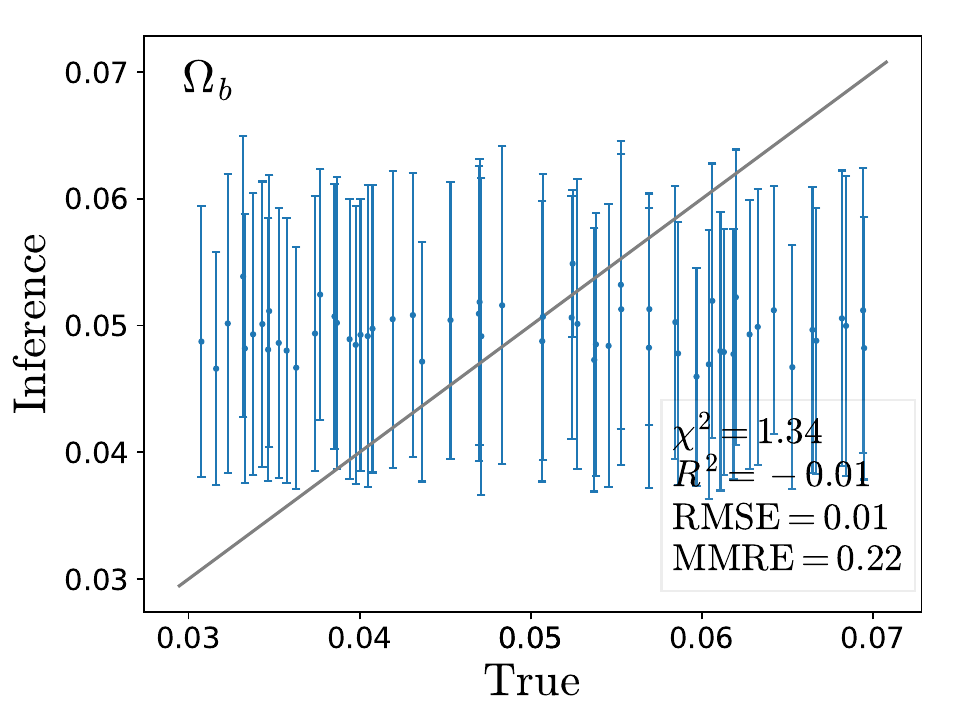}
    \includegraphics[width=.333\textwidth]{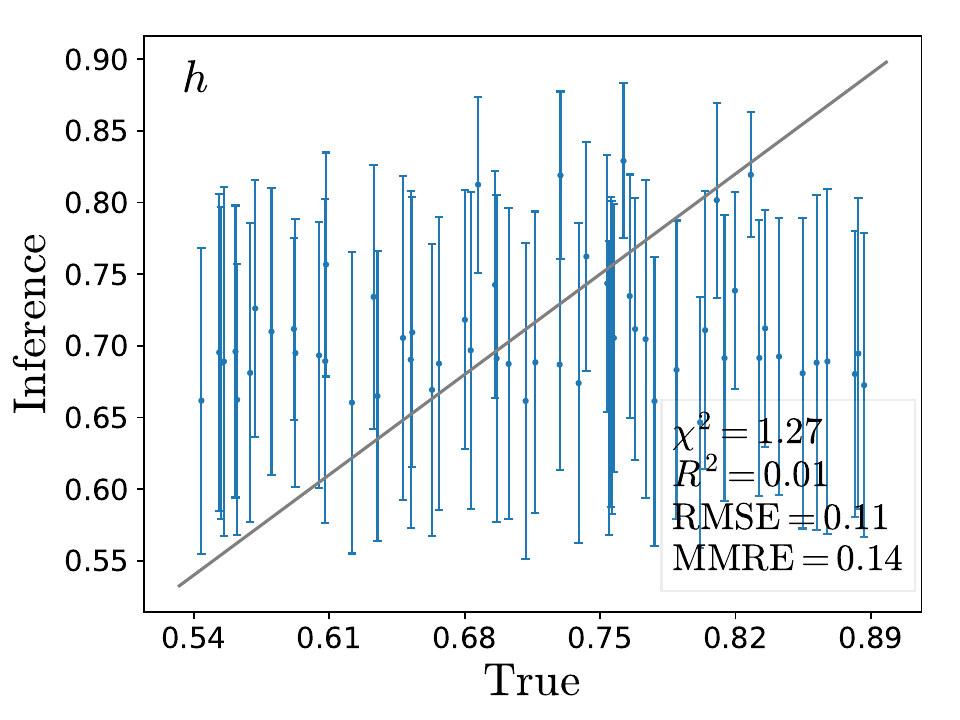}

    \caption{Plots representing the results for  the unconstrained parameters using deep sets with void radius, ellipticity, and density contrast all together.}
\end{figure*}

\FloatBarrier
\bibliography{paper}{}
\bibliographystyle{aasjournal}
\end{CJK*}
\end{document}